\documentclass[a4paper]{jpconf}
\usepackage{citesort}
\usepackage{graphicx}
\usepackage{epstopdf}
\usepackage{amsmath}
\begin{document}
\title{Geant4 based simulations for novel neutron detector development}

\author{T Kittelmann$^1$, I Stefanescu$^2$, K Kanaki$^1$, M Boin$^3$, R Hall-Wilton$^1$ and K Zeitelhack$^2$}

\address{$^1$ European Spallation Source ESS AB}
\address{$^2$ Heinz Maier-Leibnitz Zentrum, FRM-II, Technische Universit{\"a}t M{\"u}nchen}
\address{$^3$ Helmholtz-Zentrum Berlin f{\"u}r Materialien und Energie}

\ead{thomas.kittelmann@esss.se}

\begin{abstract}
A \texttt{Geant4}-based \texttt{Python}/\texttt{C++} simulation and coding
framework, which has been developed and used in order to aid the R\&D efforts
for thermal neutron detectors at neutron scattering facilities, is
described. Built upon configurable geometry and generator modules, it integrates
a general purpose object oriented output file format with meta-data, developed
in order to facilitate a faster turn-around time when setting up and analysing
simulations.  Also discussed are the extensions to \texttt{Geant4} which have
been implemented in order to include the effects of low-energy phenomena such as
Bragg diffraction in the polycrystalline support materials of the
detector. Finally, an example application of the framework is briefly shown.
\end{abstract}

\section{Introduction}

The construction of the European Spallation Source\cite{ess,esstdr}, which will
become the world's most powerful source of thermal neutrons,
is about to begin in Sweden, breaking ground in 2014 and coming online towards
the end of the decade. Currently 22 scattering instruments are planned as the
baseline suite at the facility, and a crucial part of each such beam-line will
be the detector at which neutrons are detected after undergoing scattering in a
given sample under study.

Detection of neutrons with sub-eV kinetic energies must necessarily proceed
through destructive nuclear processes in which energetic secondaries are
released. Only a few stable isotopes such as $^3$He, $^{10}$B, $^6$Li,
$^{157}$Gd and $^{235}$U have significant relevant cross sections for such
interactions, and detector systems must contain such materials as well as
incorporate capabilities to detect the resulting secondaries. The dominant
detector choice has so far been gaseous $^3$He detectors, based on the high
cross section process
$\text{n}+^{3\!\!}\text{He}\to^{3\!\!\!}\text{H}+p$. However, due to increased
demand and decreased supply, $^3$He will be unavailable in the future for all
but the smallest detectors\cite{he3crisis1,he3crisis2}. Thus, an extensive
international R\&D programme is currently under way\cite{icnd_website} in order
to develop efficient and cost-effective detectors based on other isotopes.

\begin{figure}[ht]
\begin{minipage}{0.505\textwidth}
\includegraphics[height=\textwidth,angle=270]{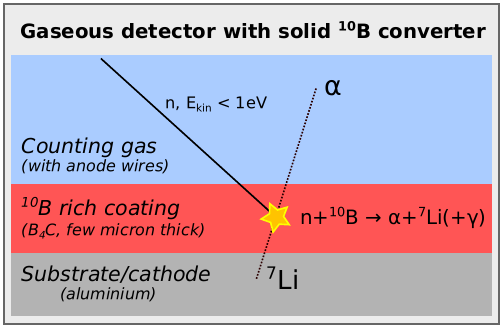}
\caption{\label{fig:boron_principle}Principle of a $^{10}$B detector.}
\end{minipage}\hspace{0.085\textwidth}%
\begin{minipage}{0.41\textwidth}
\includegraphics[height=\textwidth,angle=270]{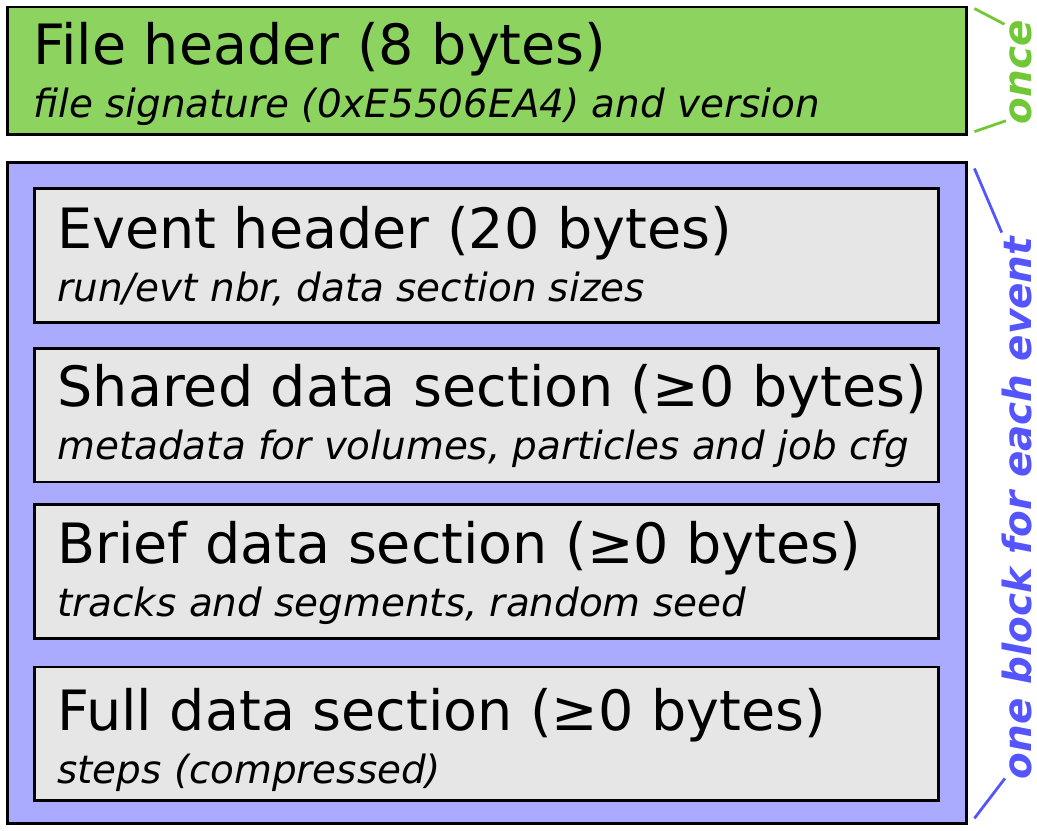}
\caption{\label{fig:griff_disklayout}Disk layout of \texttt{GRIFF} files.}
\end{minipage}
\end{figure}

A promising alternative is gaseous detectors surrounded by solid converters in
the form of thin films of $^{10}$B-enriched boron
carbide\cite{b4cfilms_carina,Andersen2013116}. The basic principle of a
successful detection event in the latter is illustrated in figure
\ref{fig:boron_principle}: after conversion
one of the released ions travel into the instrumented counting gas where it is
detected like any energetic charged particle. Such a detector has an inherent
high-rate capability and, due to the high amount of energy released in the
reaction and the implied large signals, the possibility for very high
suppression of gamma backgrounds which can otherwise be an issue at neutron
instruments. Additionally it is a relatively cheap technology, allowing for
large detector coverage when needed. However, high conversion efficiencies
requires the neutron to traverse several tens of microns of converter, while the
resulting $\alpha$ and Li ions only have a reach in solids of a few
microns. Thus, to obtain high detection efficiencies, one must either use many
independent layers of gas-facing converters, keep the angle of incidence of the
neutron on the converter as low as possible, or a combination of the two. But
such configurations inevitably include significant amounts of substrate and
support material in the path of the neutron, and although materials with a low
scattering cross section such as aluminium are chosen, the scattering herein can
still become a concern. All in all, it is highly non-trivial to access the
overall performance of a complete detector system, implying the need for
accurate full-scale simulations, such as with the framework presented here,
built upon \texttt{Geant4}\cite{geant4a,geant4b}. Whilst this framework has been
implemented specifically in order to allow for efficient and flexible
investigations of neutron detector designs, it has potential for wider
applications in neutron scattering, and in other disciplines, and is intended to
make relevant parts of it available to the wider community.

\section{General coding framework}

All code is developed in a custom multi-user coding framework where software is
kept in inter-dependant logical units ("packages"), providing libraries,
applications, scripts, (small) data files, and compiled or pure \texttt{Python}
modules\footnote{The main code branch is kept in a Mercurial repository at the
  Data Management and Software Centre at ESS/Copenhagen and supported platforms
  are currently 32 or 64 bit \texttt{Linux} or \texttt{OSX} with \texttt{GCC} or
  \texttt{LLVM/Clang}.}.

Supported languages are \texttt{C++}, \texttt{Python}, \texttt{C} and
\texttt{Fortran} with \texttt{Boost} \texttt{C++/Python} bindings.  The
configuration and detection of external dependencies is based on \texttt{CMake}
scripts, but most developers will use a single command-line tool to easily and
automatically (re-)configure, build and launch unit-tests. In addition, apart
from adding a single line to specify the inter-package dependencies (and
optionally dependencies on external software), the developer will never have to
edit any configuration files when working on a package. Instead, the names of
sub-directories in which files are placed inside a package follow certain
conventions, interpreted by the build system. As an example, placing one or more
files with extensions \texttt{.hh} and \texttt{.cc} inside a subdirectory named
\texttt{app{\textunderscore}myprog/} will result in them being compiled as
\texttt{C++} into an application named \texttt{myprog}. If the package has a
(direct or inherited) dependency on \texttt{Geant4}, the application will be
compiled and linked accordingly. Unit testing is an integral part of the
framework, and any script or application whose name starts with \texttt{test}
will be taken as a unit test.

An essential feature of the framework, is the ability to gracefully handle
missing external dependencies. While \texttt{CMake}, \texttt{Boost},
\texttt{Python} and a \texttt{C}/\texttt{C++} compiler are always required,
other dependencies such as \texttt{Geant4} are optional.
If missing, packages requiring them will be gracefully disabled, and the rest of
the code built as normal. This feature greatly reduces the installation
requirements of developers only interested in a sub-set of features, and equally
means that developers are not unduly prevented from adding new dependencies.

\section{Geant4 Simulation framework}

Embedded in the coding framework is a \texttt{Geant4}-based simulation framework
in which geometry and generator implementations are encapsulated as configurable
modules, normally instantiated and attached to the framework in a short
\texttt{Python} script. Despite typically being just a handful of lines long,
such a script is immediately a fully-featured application which one can use to
query, control and launch all relevant aspects of the chosen simulation setup
from either the command line or by adding additional calls in the script
itself. Notably, the configurability relating to geometry extends to materials
as well, which makes investigations into the role of different materials as easy
as changing a parameter at the command line (this can also be used to enable the
custom materials with crystal diffraction discussed in section
\ref{sec:nxslibintegration}). In addition to sheer convenience, the complete
configurability from either \texttt{Python} or the command line facilitates easy
scanning of parameter space when optimising proposed detector layouts.

By default, the framework is set up to write out results in the \texttt{GRIFF}
format presented in section \ref{sec:griff}. It furthermore enables
multiprocessing, handles random number streams, \texttt{Geant4} physics lists
and thresholds, and allows for the launch of a graphical viewer to inspect the
geometry and simulation results (either using the one shipped with
\texttt{Geant4} or a custom \texttt{OpenSceneGraph} based one which is currently
being developed). Finally, the framework provides a number of other useful
features, such as providing a simple switch to enable cross sections for all
active physics processes to be dumped to a separate file.

\subsection{GRIFF file format and analysis framework}\label{sec:griff}

A general purpose object oriented output file format, \texttt{GRIFF}, with
meta-data has been implemented, in order to facilitate a faster turn-around time
when setting up and analysing simulations, as well as allowing more complex and
detailed whole-event analyses. It is optimised for easy, fast and reliable
analyses from either \texttt{C++} or \texttt{Python} of low-multiplicity
physics, but supports filtering for scenarios involving higher multiplicities or
statistics.

\begin{figure}
\begin{center}
\includegraphics[height=0.75\textwidth,angle=270]{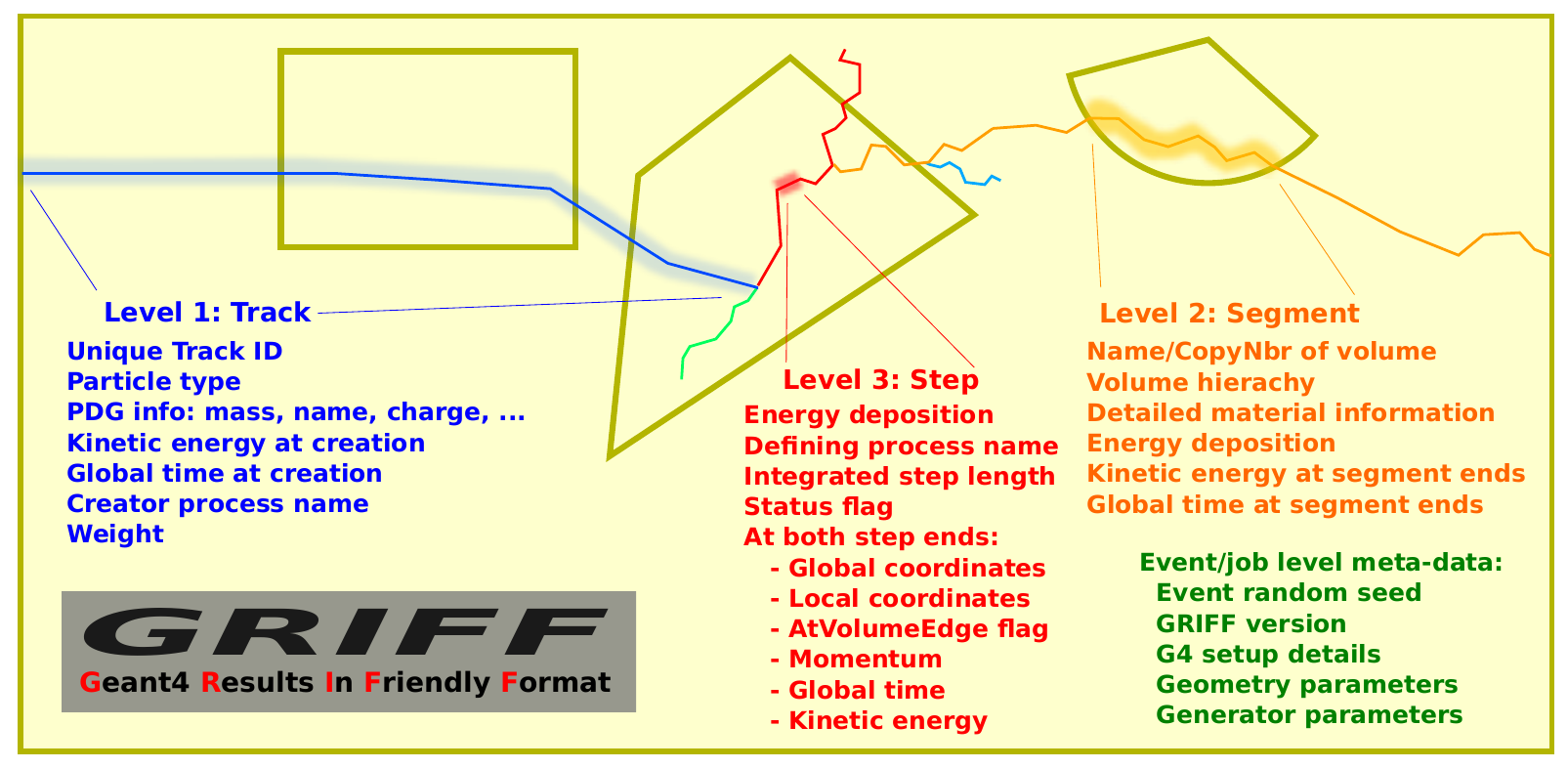}
\end{center}
\caption{\label{fig:griffuserview}Information available in a \texttt{GRIFF} file.}
\end{figure}

Installed through the \texttt{Geant4} stepping action hook, the entire event
will by default be written to the file, including information normally available
to traditional \texttt{Geant4} in-job analyses such as information about volumes
and particle data. The file can thus be opened and read without a
\texttt{Geant4} installation, and provides the user with an object oriented
access to information from the entire event as illustrated in figure
\ref{fig:griffuserview}. Easy navigation between objects (e.g. from a track to
its daughter tracks or constituent steps) is naturally included. As illustrated
in the figure, one additional layer is interspersed between the track and step
layers: series of consecutive steps of a track within a given volume are grouped
together in {\em segments}. Not only does this facilitate a more efficient data
layout, it also is a highly useful concept at the analysis level.

If necessary, \texttt{GRIFF} supports two means of filtering output. Due to the
contained coordinates, step data requires the most disk space, and the primary
global filtering method is therefore to reduce the amount of steps written
out. Either by completely omitting step data, or by combining all steps on a
segment into one meta-step. Alternatively, users can register custom
step-filters, thereby gaining complete control of the stored output.

Hidden from the user, the actual on-disk layout of \texttt{GRIFF} files is
illustrated in figure \ref{fig:griff_disklayout}: After a short file header, one
event block is appended for each event. Data inside the block is kept in three
sections containing shared, brief and full data respectively. Data unique to
individual tracks, segments and steps is kept in the two latter, while as the
name implies, common data relevant across events is kept in the shared data
section. This includes any strings and metadata relating to volumes, particles
and job configuration, which can then be referenced economically through simple
indices in the other sections (in current and following events). This means that
in order to load the $N$th event, the shared data sections of events 1 through
$N$ must have been loaded. Such a layout was chosen to enable direct streaming
to disk without the need for additional post-processing, but typically only the
first few events in a file will contain shared data.

The actual loading of an event is a highly efficient operation: The brief data
section is loaded into memory and wrapped with pre-allocated thin track and
segment classes. Data deserialisation is granular and happens only on demand
when a particular property of an object is queried, and step objects are only
created for a particular segment if needed.

\subsection{Extending Geant4 with NXSLib}\label{sec:nxslibintegration}

\begin{figure}[ht]
\begin{minipage}{0.48\textwidth}
\includegraphics[width=\textwidth]{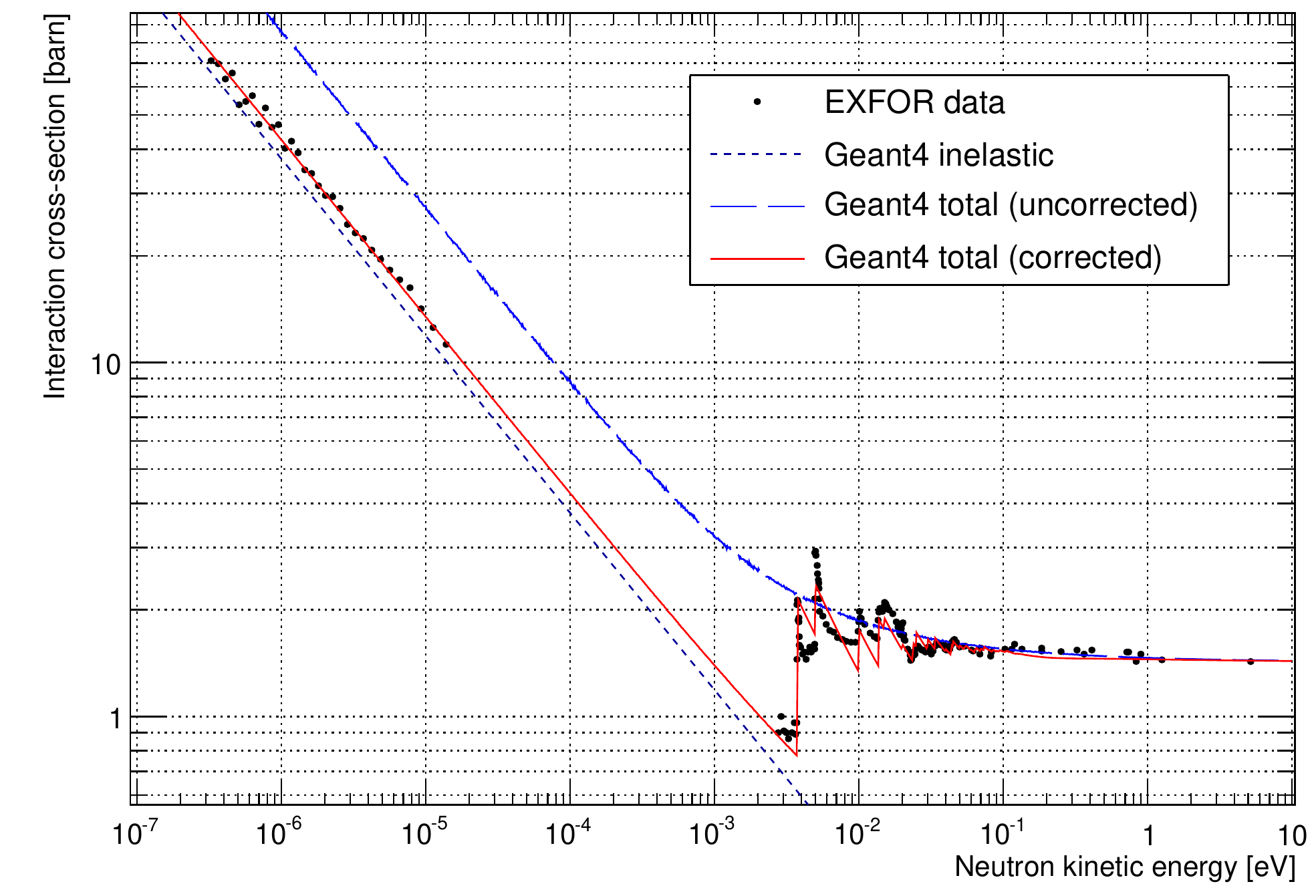}
\caption{\label{fig:nxslib_xsects}Original and \texttt{NXSLib}-corrected
  cross sections for neutron interactions in \texttt{Geant4} compared to
  data\cite{exfordata_alu_for_nxslibcomp}.}
\end{minipage}\hspace{0.04\textwidth}%
\begin{minipage}{0.48\textwidth}
\includegraphics[width=\textwidth]{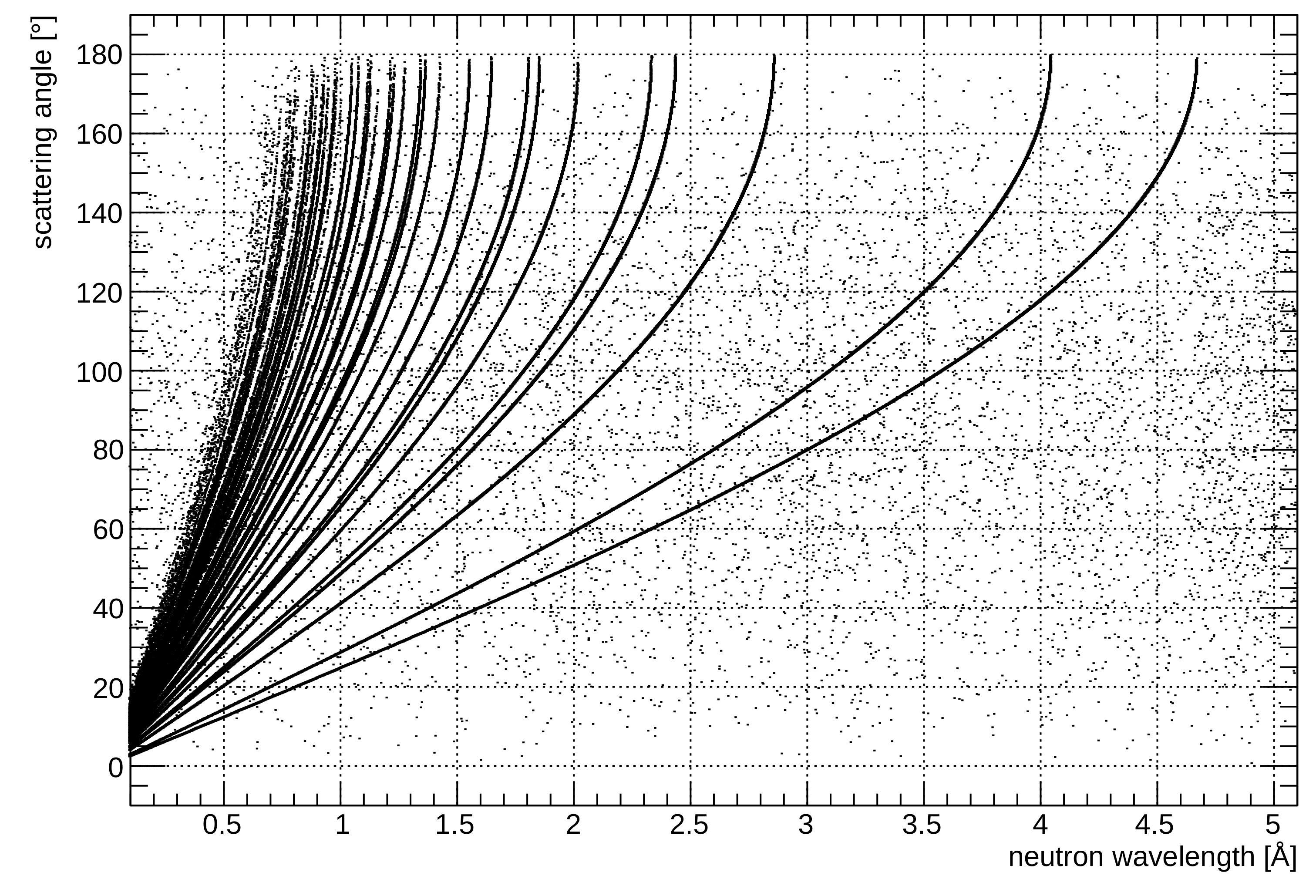}
\caption{\label{fig:nxslib_scatterangles}Scattering angles in aluminium provided
  through \texttt{NXSLib}. A similar plot based on uncorrected \texttt{Geant4}
  is featureless.}
\end{minipage}
\end{figure}

For neutron wavelengths of $\mathcal{O}(1\texttt{\AA})$, coherent scattering in
the form of Bragg diffraction dominates in polycrystalline materials such as
aluminium. This important effect is, however, not present in \texttt{Geant4} out
of the box, and has been included through integration with the polycrystal
library \texttt{NXSLib}\cite{nxslib1,nxslib2,nxslib3}. Figures
\ref{fig:nxslib_xsects} and \ref{fig:nxslib_scatterangles} show the important
effect of this correction on the relevant neutron scattering cross sections in
aluminium\footnote{Where relevant figures were created using
  \texttt{Geant4.9.6.p02} and the
  \texttt{QGSP{\textunderscore}BIC{\textunderscore}HP} physics list.}. The
\texttt{NXSLib}--\texttt{Geant4} integration augments the existing rich
capabilities of \texttt{Geant4} to become a complete tool for investigations of
a multitude of phenomena at neutron scattering facilities, and will be described
in more detail in a dedicated future publication.

\section{Example application: novel SANS detector geometry}

\begin{figure}[ht]
\begin{minipage}{0.48\textwidth}
\includegraphics[width=\textwidth]{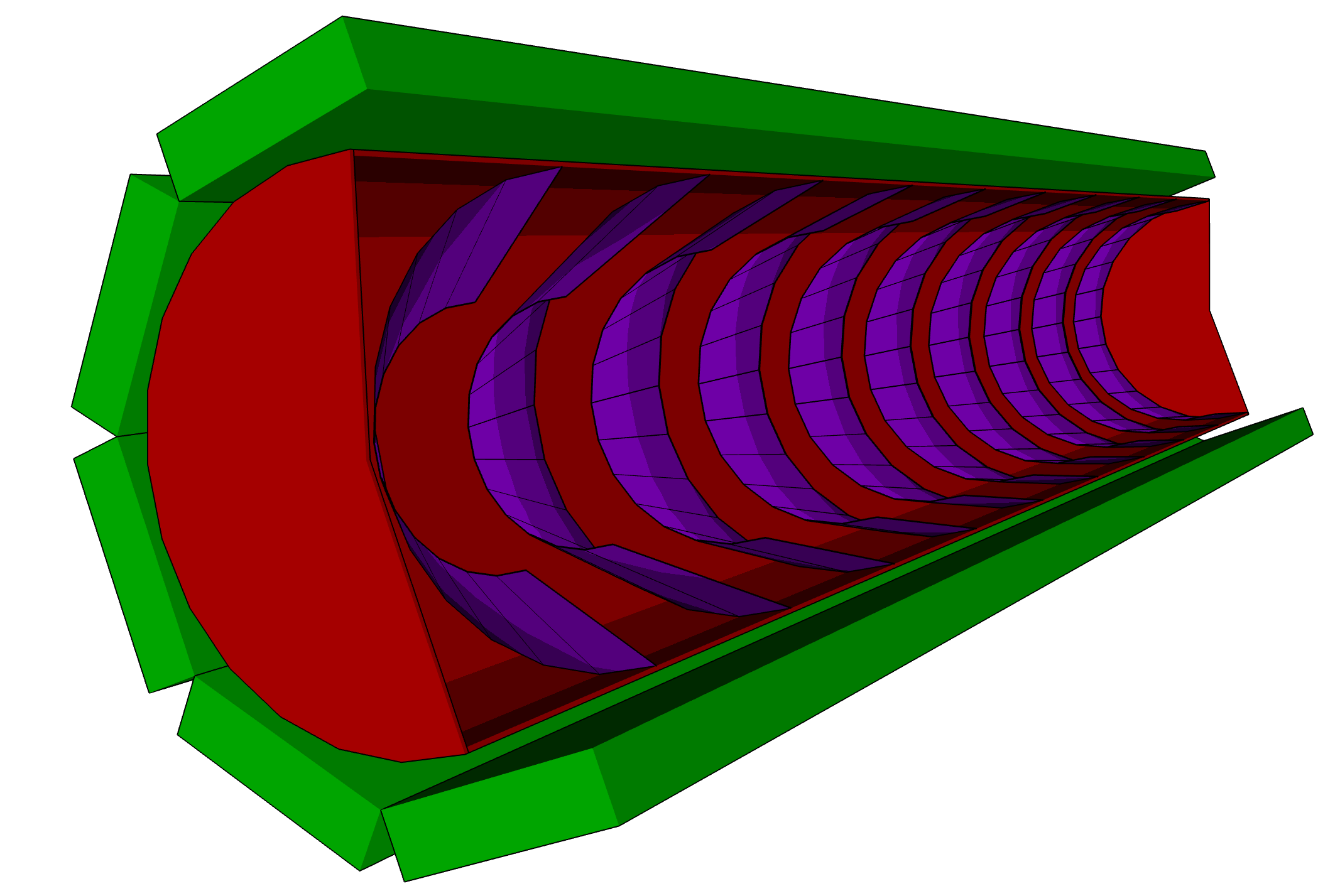}
\caption{\label{fig:sanstube3d}Proposed neutron detector with the sample located
  at the entrance of the vacuum vessel (red), with pointing absorbers to limit
  background scattering (blue/purple) and actual detector boxes (green).}
\end{minipage}\hspace{0.04\textwidth}%
\begin{minipage}{0.48\textwidth}
\includegraphics[width=\textwidth]{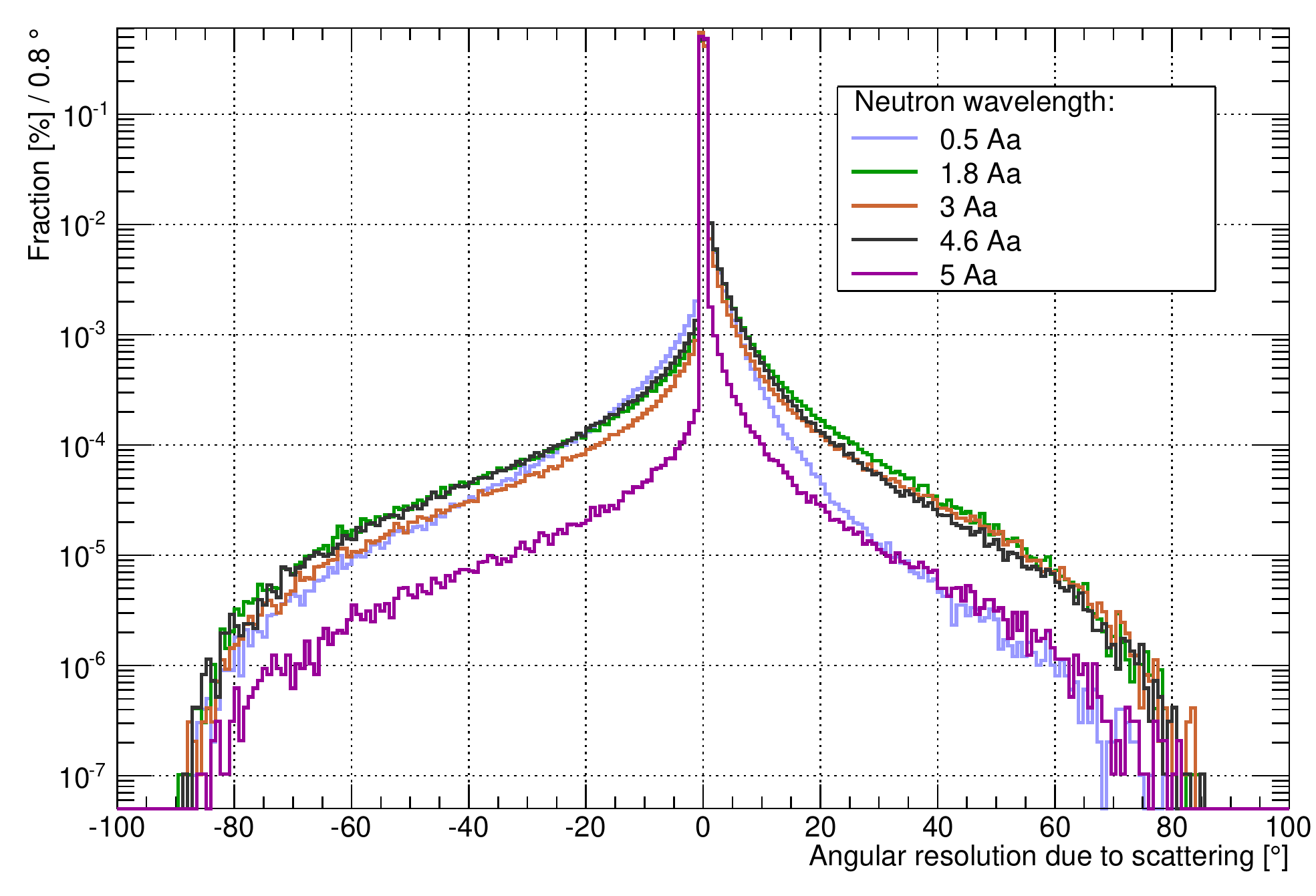}
\caption{\label{fig:sanstube_deltatheta}Difference between measured and actual
  neutron direction for different wavelengths.}
\end{minipage}
\end{figure}

An illustrative application of the framework is in the development of a novel
new detector geometry for a small angle neutron scattering
instrument\cite{sans_detgeom}, visualised in figure
\ref{fig:sanstube3d}. Diffractive scattering in support materials will be a
major challenge and simulations are essential to develop the final design of
such a detector. An
example performance plot resulting from such studies is shown in figure
\ref{fig:sanstube_deltatheta}.

\section*{References}
\bibliography{refs_thki}

\end{document}